# Studies on Size Dependent Structures and Optical Properties of CdSeS Clusters

Meenakshi Rana and Papia Chowdhury*

Department of Physics and Materials Science & Engineering,
Jaypee Institute of Information Technology, Noida 201307, Uttar Pradesh, India

**Abstract**

The size-dependent structures and optical properties of CdSeS nanoclusters in water medium are investigated. The stability of different size-dependent $Cd_nSe_mS_p$ nanoclusters (up to $n=6$) is studied using density functional theory/time-dependent density functional theory (DFT/TDDFT). The computed results for ground ($S_0$) and excited ($S_1$, $S_2$, $S_3$) states are experimentally verified through UV-Vis spectroscopy. Computed ab initio results suggest that CdSeS clusters are significantly more hyperpolarizable compared to CdX (X= S, Se, Te) clusters. Structure dependent response properties are also observed, especially for $n≥3$. Larger hyperpolarizabilities ($\beta$ and $\gamma$), charge variation and orbital analysis establish $Cd_4Se_mS_p$ clusters, as nonlinear optically active quantum dots (QDs).

**Keywords:** CdSeS nanocluster; Absorbance; Fluorescence; Hyperpolarizability.

# 1. Introduction

Over the last few years, a significant amount of research works has been reported on different confined nanometric systems such as quantum wells, quantum wires and quantum dots (QDs), etc. [1-3]. However, semiconductor QDs have become a particular point of interest in recent years due to their critically size-dependent optical properties [4]. QDs are a different class of quantum-confined semiconductor nanocrystals whose radii are smaller than bulk exciton bohr diameter (1–10 nm) [5]. Confinement results in the discretization of the electronic energy levels, in contrast to the continuous bands [6] observed in usual semiconductors. In quantum confined regime, the control of particle sizes allows the band gap to be "tuned" to give the desired electronic and optical properties such as broad excitation spectra, tunable and symmetric emission spectra with a narrow bandwidth [7,8]. QDs are also exciting new raw materials for photonics and quantum information processing because their properties can be tailored to a wide extent. However, the main attraction of QDs lies in their high optical extinction coefficients [9], tunable sharp emission profile [10], carrier multiplication ability [11,12] and variable nonlinear properties [13,14], etc., which can be tuned by adjusting size and shape. Due to their numerous fascinating properties, QDs have attracted considerable attention in many applications, such as novel luminescent sensors [15-17], quantum computing [18], solar cell [19], nano scale display devices [20], QD lasers [21], white light-emitting diode (LEDs) [22] etc.

In recent years, researchers have been focusing on group II–VI semiconductor materials like CdSe [23], CdS [24,25], ZnS [26], ZnSe [27], etc., as core type QDs. Most of these investigations were focused on the synthesis and structure determination of QDs along with the evaluation of some concomitant physical features. Chelikowsky et al. in 2001 have been done a theoretical study on structural and electronic properties of $Cd_nS_n$ and $Cd_nSe_n$ clusters with upto $n$=8 [28]. After that there are alarge number of research works have been reported on Cd based QD clusters, to explore the different properties such as structure, electronic, polarizability, hyperpolarizabilities [29,30]. Recently in 2010 Gutsev et.al. investigate the geometrical structural and electronic properties of GaAs neutral clusters by using generalized gradient approximation [31]. QD-related science has made a significant and fruitful progress in the field of nonlinear optics (NLO) also. NLO processes are interesting since they are associated with some modern opto-technological applications [32,33]. However, studies on most of the optical properties are mainly limited to $(CdX)_n$ [X: S, Se, Te] type bare and core/shell QDs, such as ZnSe/ZnS and CdSe/ZnSe QDs [34,35]. Very little attention has been given to the alloyed $(Cd_nX_mY_p)$ QDs like CdHgTe, CdSeS, etc. as core material. CdS (band gap $(\Delta\varepsilon)$=2.48 eV, lattice constant (a)= 6.050 Å) and CdSe ($\Delta\varepsilon$=1.73 eV, a=5.835 Å) promotes the formation of CdSeS alloyed QDs synthetically very easily due to very small lattice mismatch between them [35]. Different types of structural models are available of CdSeS ternary alloyed QDs, such as homogeneous alloyed dots and gradient alloyed dots. Different structural models influences the electronic and optical properties of QDs [36]. In the present work, our interest is focused on a specific hybrid nano sized cluster $Cd_nX_mY_p$ [X: Se, Y: S]. Though Cadmium (Cd) is a hazardous to environment, but CdSeS show no obvious toxicity since protective CdS shells endow high stability and low cytotoxicity. CdSeS exhibits good optical properties with high fluorescence quantum yield (QYs) (up to 25%) and a narrow Full width at half maximum (FWHM) of 28 nm [35]. For $Cd_nSe_mS_p$ formation, the growth of the CdS layer used to appear on the surface of CdSe core. CdS layer behaves like shell to CdSe core which effectively reduce the nonradiative trap and increases the QY. A large change in optical properties of CdSeS QD is observed due to the variation in its compositions and so the response properties of such clusters are characterized by their large variations in NLO coefficients. A number of optical nonlinearities also allow CdSeS QDs to be used as excellent candidate for making of fluorescence marker, optical switches [35,36]. All the above mentioned characteristic properties of the novel CdSeS QDs made us curious to investigate about its exact geometrical structures and optical properties. In this letter we have presented a detailed systematic computational and experimental investigations on the geometrical structure, linear and nonlinear optical properties of stoichiometric small $Cd_nSe_mS_p$.

## 2. Materials and Methods

## 2.1 Materials

Cadmium chloride hemipentahydrate ($CdCl_2 \cdot 2.5H_2O$), zinc acetate dihydrate [$Zn(Ac)_2 \cdot 2H_2O$], anhydrous sodium sulfite ($Na_2SO_3$), 3-mercaptopropionic acid (MPA), selenium powder (Se), sodium sulfide ($Na_2S$), reagent grade ethanol and isopropanol were purchased from Sigma Aldrich for the preparation of CdSeS QD. A microwave digestion system was used for the synthesis of QDs with different controllable temperature and pressure units. MPA act as a stabilizing agent. $Na_2SeSO_3$ solution with 0.30 molL$^{-1}$ concentration was prepared as report by Hankare et. al. [35] with Se powder (2.3670 gm, 0.030 mol), $Na_2SO_3$ (9.453 gm, 0.075 mol) and double distilled water (80 mL) and used as stock solution as a source of Se. A one step procedure is adopted for the preparation of CdSeS QDs reported by H. J. Zhan et.al.[37]. The CdSeS precursor solution was prepared by adding $Na_2SeSO_3$ solution to $CdCl_2$ solution (pH 9.0) containing MPA. 8:20:1 molar ratio of Cd: MPA: Se was used for the QD synthesis. We can also feed different molar ratios of Cd:Se:S in the preparation of CdSeS ternary QDs such as (2:0.6:1.4), (3.5:0.6:1.4), (2:1:1), (3.5:1:1), (3.5:1.4:0.6) etc. [38]. Sulfide ions (S) released from MPA at a high temperature (130 °C). The resultant precurse solution of CdSe was placed in a teflon coated inner vessel under microwave digestion furnace for the formation of CdSeS alloyed QDs. For perfect optimization with maximum 30 % photo luminescence quantum yield (PLQY) of CdSeS QDs, 40 min of microwave irradiation was applied at 130°C. Teflon inner vessel provides security in the reactions demanding higher temperature and pressure. Finally, the CdSeS QDs were naturally cooled at room temperature and purified with by the addition 2-propanol [37]. After purification, the clean QD solution were adjusted to a concentration with an absorption value of ~0.1 [37].

## 2.2 Experimental Methods

The UV-Vis absorption spectra were recorded at 300 Kelvin by a Perkin Elmer spectrophotometer (model Lambda-35) with a varying slit width in the range 190-900 nm. All luminescence measurements were made with a Perkin Elmer spectrophotometer (Model Fluorescence-55) with a varying slit width (Excitation slit = 10.0 nm and Emission slit = 5 nm) ranging from 200-900 nm. The Model LS 55 series uses a pulsed Xenon lamp as a source of excitation. Deionized water (Milipore) was used for measuring absorption and emission spectra. All optical measurements were performed at room temperature under ambient condition.

## 2.3 Computational methods

All the quantum chemical computations, including the ground state ($S_0$) and excited state ($S_1$) optimizations and energy calculations are carried out using density functional theory (DFT) [39, 40] by Gaussian 09 [41]. We have optimized the most stable geometry of CdSeS cluster with coupled-cluster singles and doubles method (CCSD). The ground state energy obtained at the DFT level of theory was -25350.3045 a.u., and the value obtained at the CCSD level was - 25337.1337 a.u. The predicted magnitude of enegy by DFT is much smaller than the CCSD one. So we have used the DFT method for further computation. The ground state geometry of $Cd_nSe_mS_p$ clusters were optimized with symmetry constraints with Becke's three parameter hybrid exchange function and Lee–Yang–Parr gradient corrected correlation functional (B3LYP hybrid functional) [42] by using a 3-21G* basis set.. The stable, optimized geometries are subsequently used as starting structures for all other calculations. Frequency calculations were performed on optimized structures. Frequency calculations serve to compute thermodynamic quantity of interest such as entropy (S) of the system. S is calculated with S=(G-H)/T, where H is the enthalpy, G is Gibbs free energy and T is the temperature for probe system. All the computations for the system are carried out at 1 atmosphere pressure and temperature T= 298.15 K. The vertical excitation energies $S_0 \rightarrow S_1$, $S_2$ and $S_3$ states of molecules in water medium were performed using TDDFT. The excited state geometries were optimized with TDDFT by using B3LYP hybrid functional with the same basis set. To study the emissive properties, optimized geometry of a specific excited state is obtained by the guess geometry of the $S_0$ state [43]. Oscillator strengths and dipole moments were deduced from the dipole transition matrix elements. Macroscopic linear and nonlinear optical response of nanocluster is connected with the microscopic polarizability and hyper polarizabilities. The NLO properties of CdSeS clusters such as, dipole moment ($\mu$), mean polarizability ($\alpha$) and hyperpolarizabilities ($\beta$ and $\gamma$) are computed with static applied electric field strength for specific medium. For water medium we have used the value of static dielectric constant as 78.3 and dynamic dielectric constant as 1.77. The $\alpha$, $\beta$, anisotropy of polarizability ($\Delta\alpha$), and second order hyperpolarizability ($\gamma$) of CdSeS clusters in terms of x, y, z components. $\mu$, $\alpha$, $\beta$ and $\gamma$ can be defined and computed by classic Taylor series expansion of the perturbed energy of a cluster in the presence of a weak uniform external static field [44].

The value of $\alpha$, $\Delta\alpha$, $\beta$ and $\gamma$ can be calculated by using Eqs. (1) – (4)

$$\alpha = \frac{(\alpha_{XX} + \alpha_{YY} + \alpha_{YY})}{3}, \ldots\ldots\ldots\ldots\ldots\ldots\ldots(1)$$

$$\Delta\alpha = 2^{-\frac{1}{2}}[(\alpha_{XX} - \alpha_{YY})^2 + (\alpha_{YY} - \alpha_{ZZ})^2 + (\alpha_{ZZ} - \alpha_{XX})^2 + 6(\alpha_{XY}^2 + \alpha_{YZ}^2 + \alpha_{ZX}^2)]^{\frac{1}{2}}, \ldots\ldots\ldots(2)$$

where $α_{XX}$, $α_{YY}$, $α_{ZZ}$, $α_{XY}$, $α_{YZ}$ and $α_{ZX}$ are the tensor components of polarizability.

$$\beta = [(\beta_{XXX} + \beta_{XYY} + \beta_{XZZ})^2 + (\beta_{YZZ} + \beta_{YYY} + \beta_{YXX})^2 + (\beta_{ZXX} + \beta_{ZYY} + \beta_{ZZZ})^2]^{\frac{1}{2}} \quad \ldots\ldots(3)$$

where $β_{XXX}$, $β_{YYY}$, $β_{ZZZ}$, $β_{XYY}$, $β_{XZZ}$, $β_{YZZ}$, $β_{YXX}$, $β_{ZXX}$ and $β_{ZYY}$ are the tensor components of first order hyperpolarizability.

The equation for average $γ$ is given by:

$$\gamma = \frac{1}{5}(\gamma_{XXXX} + \gamma_{YYYY} + \gamma_{ZZZZ} + 2\gamma_{XXYY} + 2\gamma_{XXZZ} + 2\gamma_{YYZZ}), \quad \ldots\ldots(4)$$

where $γ_{XXXX}$, $γ_{YYYY}$, $γ_{ZZZZ}$, $γ_{XXYY}$, $γ_{XXZZ}$ and $γ_{YYZZ}$ are the tensor components of second order hyperpolarizability.

## 3. Results and Discussion

### 3.1 Geometrical structure

Clusters are the assembly of atoms. It can be made up of semiconductors, insulators or metals. Until now there a number of work have been reported on the geometrical structure of metal clusters such as boron doped silver [45], iron oxide [46] and palladium [47], vanadium phthalocyaninato based clusters [48], semiconductor clusters like niobium-doped silicon clusters [49], nanoscale gold clusters [50] etc. Nanoscale gold nanocluster and semiconductor nanocluster play an important role in many fields such as colloidal chemistry, medical science, and catalysis [51]. Semiconductor nanocluster have been received special attention in the literature due to their unusual properties. Extensive studies on various CdX nanoclusters: CdSe, CdS, CdTe, etc., in gas and solution phase has been predicted in different literatures [44,52,53], but as per our knowledge the information regarding CdSeS alloyed structure in gas or solution phase is not reported anywhere.

To construct perfect shape and size, different bare structures of most stable semiconductor $Cd_nSe_mS_p$ clusters have been constructed. In the present manuscript we have constructed different $Cd_nSe_mS_p$ clusters with the fixed value of Cd atoms by varying the number of Se and S atoms in a structure. So we have used CdSeS monomer and $Cd_nSe_mS_p$ (up to $n=6$) cluster structures, with the condition $n=m+p$ as our probe system. DFT computational approach has been previously used to study the various low-lying sulphur clusters [54], boron clusters [45], neutral and anionic sodium clusters [55] and niobium-doped silicon clusters [49]. In this work, different $Cd_nSe_mS_p$ clusters with their planar and nonplanar structures have been optimized with DFT by 3-21G basis set [44,52].

In solution phase, $(CdSe)_n$ and $(CdS)_n$ clusters show both planar, ring and 3D structures [53,56], structures $Cd_nSe_mS_p$ clusters also remain consistent with previous studies [56]. Scheme 1(A-F) illustrates the ground state QD structures arranged in order of increasing size in terms of their point groups, bond lengths, bond angles and natural atomic charges. For $Cd_nSe_mS_p$ formation CdS layer behaves like shell to CdSe core [37]. Optimized results show that different possible planar and nonplanar $Cd_nSe_mS_p$ clusters are $Cd_1Se_1S_1$, $Cd_2Se_1S_1$, $Cd_3Se_2S_1$, $Cd_3Se_1S_2$, $Cd_4Se_2S_2$, $Cd_4Se_1S_3$, $Cd_4Se_3S_1$, $Cd_5Se_3S_2$, $Cd_5Se_2S_3$, $Cd_5Se_4S_1$, $Cd_5Se_1S_4$, $Cd_6Se_3S_3$, $Cd_6Se_2S_4$, $Cd_6Se_4S_2$, $Cd_6Se_1S_5$, and $Cd_6Se_5S_1$. The lowest energy structure of $Cd_1Se_1S_1$ clusters possess a linear structure with $C_{\infty V}$ (Scheme1 A). For $Cd_1Se_1S_1$ cluster, two linear configurations are possible which are are shown in Scheme 1A and SD 1a. Out of both linear isomers, the structure of Scheme 1A comes out to be more stable energetically with Cd–Se and Cd–S bond lengths 2.52 and 2.42 Å respectively. $Cd_1Se_1S_1$ can form in plane triangular structures also but all triangular structures were found to be unstable energetically. Some $Cd_2Se_1S_1$ cluster structures are shown in Scheme1 B and SD 2a,b,c. $Cd_2Se_1S_1$ cluster have four possible structures (Scheme1 B and SD 2a,b,c).

The lowest energy structure holds rhombic structure (Scheme1 B). The bond length between two Cd atoms is 2.65 Å and the Cd-Se, Cd-S separation is 2.83 Å, 2.47Å respectively. The two distorted trapezoidal structures SD 2a and SD 2b are almost degenerate in energy, and they have energy above the lowest energy structure. The linear $Cd_2Se_1S_1$ isomer (SD 2c) is higher in energy than the other available structures. Lowest lying isomers of $Cd_3Se_mS_p$ clusters are showm in Scheme1 C,D and SD 3a-j. The theoretical results show that the lowest-lying configurations for $Cd_3Se_mS_p$ clusters are planar. The $Cd_3Se_2S_1$ isomer Scheme1 C1 lies the lowest energy isomer. Other structures shown in SD 3a-e have their total energies are higher than the lowest-energy isomer of Scheme1 C1. Similarly, we note that when an S atom replaces one Se atom of the $Cd_3Se_2S_1$ isomers, the isomers of $Cd_3Se_1S_2$ clusters are yielded. These structures are shown in Scheme1 C2 and SD 3f-i. In the lowest energy structure of $Cd_3Se_mS_p$ clusters, three Cd atoms form a regular triangle and possess $C_1$ symmetry (Scheme1 C1,C2). It is observed that complexity in the structure of both planar and nonplanar structures increases from $n≥3$ due to possibility of more number of cluster formation for a fixed number (*n*) of Cd atom. Optimized structures suggested that the clusters prefer planar structure upto $n=3$ and three dimensional structure is preferred only on or after $n≥3$. We have checked all possible planar and 3D models for the $Cd_nSe_mS_p$ (up to $n=6$) clusters, such that the $n=m+p$, and checked their stability acoording to their energy. Scheme 1 displays the possible lowest energy clusters.

In case of $Cd_4Se_mS_p$ clusters, we have noticed that the nonplanar (3D) structure having distorted tetrahedral geometry with $C_1$ point group symmetrires exhibits higher stability compared to the planar structure (Scheme D1-D3). For the pentamer unit of $Cd_5Se_mS_p$ clusters global minima is observed with the nonplanar geometry with $C_{2V}$ point group symmetry (Scheme E1-E5). The hexamer $Cd_5Se_mS_p$ clusters with $C_1$ point group (Scheme F1-F6) are more stable than their planar counter part and have a chair like structure. Out of all possible clusters ($n=1-6$) energetically favorable planar and nonplanar structures are $Cd_1Se_1S_1$, $Cd_2Se_1S_1$, $Cd_3Se_2S_1$, $Cd_4Se_3S_1$, $Cd_4Se_3S_1$, $Cd_6Se_5S_1$. It is observed that clusters with same configuration with nonplanar distorted geometry exhibits higher stability compared to the planar cluster structures.

## 3.2 Geometrical parameters

The different geometrical and physical parameters of closely lying structures of $Cd_nSe_mS_p$ are given in Table 1, along with their energy, dipole moment, entropy, difference between highest occupied molecular orbital and lowest unoccupied molecular orbital (HOMO-LUMO) energy band gap values and point group symmetries. An estimation of ground state energies of $Cd_nSe_mS_p$ clusters also reveals that for a fixed number of Cd atoms as the number of selenium atom increases than sulpher atom the stability of the cluster structure increases. Computed results also show that the HOMO-LUMO gap ($\Delta\varepsilon$) becomes remarkably low up to $n=2$, but $\Delta\varepsilon$ becomes larger for $n≥3$. Se has atomic number 34. It has an electron configuration of 2-8-18-6 and has 6 valence electrons. S, having atomic number 16 and electron configuration is 2-8-6 with 6 valence electron. So, for fixed number of Cd atoms, the number of valence electrons is same. The Se and S atoms are comparable in size. Atomic radius of Se is 115 pm and S has 100 pm. So one can easily replace S by Se or Se by S. When the Se concentration is increased instead of S in $Cd_3Se_2S_1$, the cluster structure may be tight bounded due to larger diameter of Se than S, which justifies the stability of $Cd_3Se_2S_1$ structure. Whereas for increased S concentration in $Cd_3Se_1S_2$, the cluster structure may appear to be loosely bound due to smaller diameter of S than Se. The more tightly bound a system is, stronger the binding forces that hold the atoms together and so greater energy is required to pull them apart. The binding energy ($E_b$) of a system defines as the energy required to completely disassembling the system. So greater $E_b$ means better stability. Greater $E_b$ means lower entropy of the system and so more thermodynamic stability. The binding energy of $Cd_nSe_mS_p$ with size $n+m+p$ can be calculated by using the equation as follows:

$$E_b(Cd_nSe_mS_p) = \frac{E(Cd_nSe_mS_p) - mE(Cd) - nE(Se) - pE(S)}{n+m+p} \quad \ldots(5)$$

For $Cd_nSe_mS_p$ cluster structures, $E_b$ appears to be larger, especially when the Se concentration is higher than S. $Cd_3Se_2S_1$ shows better stability since it has larger value of $E_b$ (-2.05 eV) in comparison with $Cd_3Se_1S_2$ (-1.8 eV). The stability of $Cd_3Se_2S_1$ cluster is further verified by its lower entropy (104.13 cal/mol-kelvin) value than of $Cd_3Se_1S_2$ cluster (115.48 cal/mol-kelvin). Similarly, for available possibility of $Cd_4Se_mS_p$ clusters, energetically most stable structure is $Cd_4Se_3S_1$ which is nonplanar in nature with $C_1$ point group symmetry. $Cd_4Se_3S_1$ cluster also exhibits better stability due to its lower entropy (95.15 cal/mol) value than the other nonplanar counterparts of $n=4$ (Table 1). It also shows more stability than its planar counterpart by an amount 17.57 kcal/mol (SD 2). It is worth mentioning that in the geometry optimization of $Cd_4Se_mS_p$ we did not obtain any global minimum and so corresponding local minimum have been taken into account for the evaluation of response properties. $Cd_4Se_3S_1$ justifies its better stability in terms of energy difference ($\Delta E$=2503.35-2403.35 kcal/mol), dipole moment ($\Delta \mu$=0.03-6.54 D) and the change in entropy ($\Delta S$=12-34 cal/mol) which justifies its lower reactivity compared to its neighbhours, ($Cd_4Se_1S_3$, $Cd_4Se_2S_2$). Out of all planar and nonplanar cluster structures for $n=4$ (SD 2), nonplanar $Cd_4Se_1S_3$ has smaller $\Delta\varepsilon$, larger $\mu$ and S, which indicates this cluster has larger reactivity to interact with the neighboring environment than the others. For all available $Cd_5Se_mS_p$ structures, $Cd_5Se_4S_1$ shows better stability energetically, entropically. It also shows higher reactivity with a higher dipole moment and smaller $\Delta\varepsilon=2.25$ eV among other pentamer structures. Similar results obtained for $Cd_6Se_mS_p$ structures. $Cd_6Se_5S_1$ shows better stability and reactivity among all other $Cd_6Se_mS_p$ structures. Out of all possible $Cd_nSe_mS_p$ cluster structures $Cd_4Se_3S_1$ comes out to be most stable and reactive structure due to its most ordered structure (lowest entropy).

To identify stability and reactivity of a specific cluster structure we have calculated the second-order energy differential ($\Delta_2E$) which reflects the relative stability of a cluster with respect to its neighbours, is defined as [57].

$$\Delta_2 E(n) = E(n-1) + E(n+1) - 2E(n), \quad\quad\quad\quad\quad\quad\quad\quad\quad\quad\quad\quad\quad\quad\quad\quad\quad\quad (6)$$

where $E(n)$ is the energy of energetically stable $n^{th}$ order cluster. $\Delta_2E$ is a measure of gain in energy by the formation of cluster of size $n$ due to the cohesion of a unit to size $n$-1 or fragmentation of size $n$+1. Higher value of $\Delta_2E$ reflects the stability and lower value of $\Delta_2E$ reflects reactivity of a specific cluster. Higher value of $\Delta_2E$ also indicates thermodynamic stability of a cluster structure. Computed $\Delta_2E$ values of all combinations of energetically stable $Cd_nSe_mS_p$ clusters ($n$=2-5) are shown in SD 3 with the variation of $n$ values excluding $n$=1 since no cluster structure is possible for $n$=0.

Scheme 1: Different optimized structures of $Cd_nSe_mS_p$, ($n$=1-6: $m$=1-5: $p$=1-5) clusters.

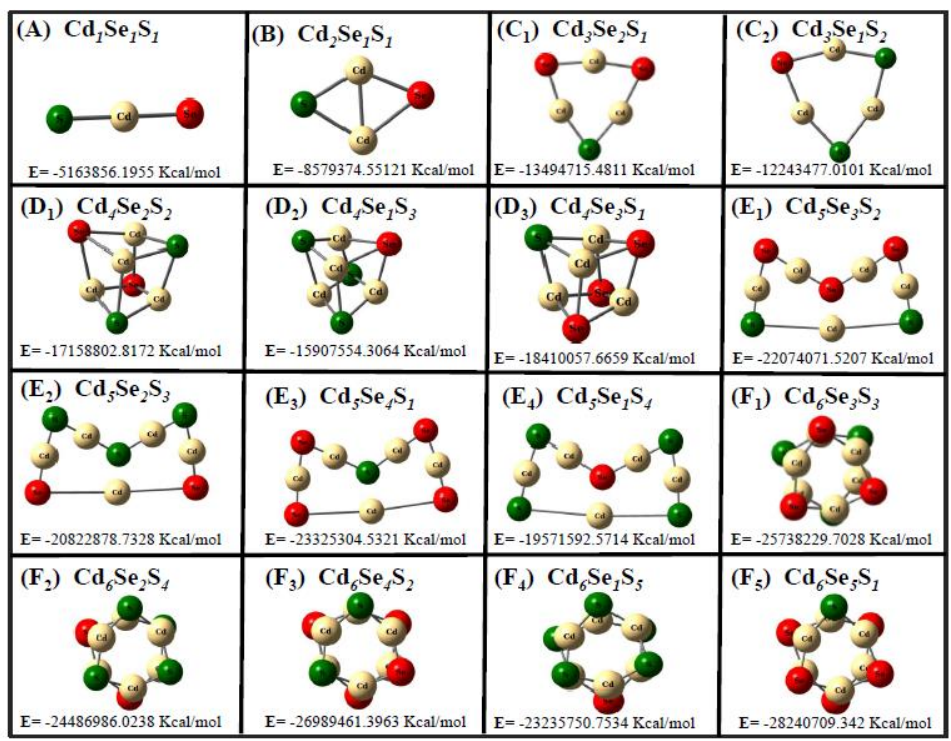

Computed $\Delta_2E(n)$ values show that the clusters with $n$=3-5 ($Cd_3Se_2S_1$= -4.13 eV, $Cd_4Se_3S_1$= -0.05 eV and $Cd_5Se_4S_1$= -6.84 eV) have higher relative stability and lower reactivity than cluster with $n$=2 ($Cd_2Se_1S_1$=-65038 eV). Interestingly for $Cd_4Se_mS_p$ clusters ($n$=4), $Cd_4Se_1S_3$ shows minimum value of $\Delta_2E$ whereas $Cd_4Se_3S_1$ shows maximum value of $\Delta_2E$ among all available $Cd_nSe_mS_p$ clusters ($n$=2-5). With maximum $\Delta_2E$ value, $Cd_4Se_3S_1$ cluster reflects its higher stability and lower reactivity, whereas with minimum $\Delta_2E$ value, $Cd_4Se_1S_3$ shows its higher reactivity and lower stability compared to other $Cd_nSe_mS_p$ clusters. High value of $\Delta_2E$ for $Cd_4Se_3S_1$ may be due to the stabilization of structure by neighbouring water environment. Similarly, for higher $n$ value, $Cd_5Se_2S_3$, $Cd_6Se_5S_1$ structures show better reactivity among other $Cd_5Se_mS_p$ and $Cd_6Se_mS_p$ clusters (not shown in SD 3).

Table 1: Different computed geometrical parameters of $Cd_nSe_mS_p$, ($n$=1-6: $m$=1-5: $p$=1-5) clusters.

| S. No. | Clusters | Symmetry | Dipole moment ($\mu$) in Debye | Energy (E) in Kcal/mol | HOMO Energy in Kcal/mol | LUMO Energy in Kcal/mol | Band Gap ($\Delta\varepsilon$) in eV | Entropy (S) in cal/mol-kelvin |
|---|---|---|---|---|---|---|---|---|
| 1. | $Cd_1Se_1S_1$ | $C_{\infty V}$ | 1.47 | -5163856.1955 | -154.6309 | -140.9700 | 0.59 | 74.87 |
| 2. | $Cd_2Se_1S_1$ | $C_s$ | 3.21 | -8579374.55121 | -132.1676 | -113.3470 | 0.81 | 85.79 |
| 3. | $Cd_3Se_2S_1$ | $C_1$ | 1.43 | -13494715.4811 | -117.8023 | -26.1357 | 3.97 | 104.13 |
| 4. | $Cd_3Se_1S_2$ | $C_1$ | 0.61 | -12243477.0104 | -118.8565 | -27.6417 | 3.95 | 115.48 |
| 5. | $Cd_4Se_2S_2$ | $C_1$ | 1.24 | -17158802.8172 | -120.6010 | -33.2642 | 3.78 | 129.97 |
| 6. | $Cd_4Se_1S_3$ | $C_1$ | 7.75 | -15907554.3064 | -117.6768 | -35.41664 | 3.56 | 122.50 |
| 7. | $Cd_4Se_3S_1$ | $C_1$ | 1.21 | -18410057.6659 | -115.4366 | -30.4279 | 3.68 | 95.15 |
| 8. | $Cd_5Se_3S_2$ | $C_{2V}$ | 2.23 | -22074071.5207 | -105.9675 | -50.16311 | 2.41 | 147.73 |
| 9. | $Cd_5Se_2S_3$ | $C_{2V}$ | 12.36 | -20822878.7328 | -113.0584 | -33.9921 | 3.42 | 147.95 |
| 10 | $Cd_5Se_4S_1$ | $C_{2V}$ | 2.45 | -23325304.5321 | -102.4723 | -50.5145 | 2.25 | 131.40 |
| 11 | $Cd_5Se_1S_4$ | $C_{2V}$ | 2.30 | -19571592.5714 | -106.6389 | -50.8973 | 2.41 | 138.05 |
| 12 | $Cd_6Se_3S_3$ | $C_1$ | 1.32 | -25738229.7028 | -120.1743 | -35.1970 | 3.68 | 168.29 |
| 13 | $Cd_6Se_2S_4$ | $C_1$ | 1.65 | -24486986.0238 | -121.5988 | -36.2198 | 3.70 | 145.98 |
| 14 | $Cd_6Se_4S_2$ | $C_1$ | 1.22 | -26989461.3963 | -120.0927 | -34.8895 | 3.69 | 149.56 |
| 15 | $Cd_6Se_1S_5$ | $C_1$ | 1.53 | -23235750.7534 | -121.6490 | -36.6465 | 3.68 | 157.76 |
| 16 | $Cd_6Se_5S_1$ | $C_1$ | 2.62 | -28240709.342 | -116.8611 | -33.9106 | 3.59 | 173.75 |

Dipole moment in Debye; Energy (E) in Kcal/mol; Band gap ($\Delta\varepsilon$) in eV; Entropy in cal/mol-kelvin.

In the present work we have also tried to characterize size dependent nonlinear optical properties of different optimized, stable cluster forms of $Cd_nSe_mS_p$ upto $n=6$ on the basis of static mean polarizability ($\alpha$), first and second order hyperpolarizabilities ($\beta$, $\gamma$) components (Table 2). We have observed an increasing tendency in value of $\alpha$ for all types of clusters except for $n=2$ [SD 4, Table 2]. However, no such trend is observed for $\Delta\alpha$. The observed values of the $\alpha$ and $\Delta\alpha$ or CdSeS clusters seem to be significantly high as compared with other known CdX clusters with X=S, Se, Te [51] respectively. We have calculated $\beta$ for all the clusters ($n$=1-6) for energetically most stable geometries and for closely lying geometries.

Table 2: Computed mean polarizability ($\alpha$), anisotropy of polarizability ($\Delta\alpha$), mean first and second order hyperpolarizability ($\beta$, $\gamma$) of $Cd_1Se_1S_1$, $Cd_2Se_1S_1$, $Cd_3Se_2S_1$, $Cd_4Se_1S_3$, $Cd_4Se_3S_1$, $Cd_5Se_4S_1$ and $Cd_6Se_5S_1$ clusters in water, KDP and urea.

| S. No. | Clusters | Mean polarizability ($\alpha$) | Anisotropy of polarizability ($\Delta\alpha$) | Mean first order hyper polarizability ($\beta$) | Mean second order hyper polarizability ($\gamma$) |
|---|---|---|---|---|---|
| 1. | $Cd_1Se_1S_1$ | 203.29 | 447.50 | 753.81 | -581.97 |
| 2. | $Cd_2Se_1S_1$ | 174.78 | 164.35 | 1929.21 | -677.22 |
| 3. | $Cd_3Se_2S_1$ | 276.51 | 567.39 | 661.54 | -1476.07 |
| 4. | $Cd_4Se_3S_1$ | 423.63 | 738.27 | 594.75 | -11416.39 |
| 5. | $Cd_4Se_1S_3$ | 385.94 | 661.24 | 3166.16 | -1845.12 |
| 6. | $Cd_5Se_4S_1$ | 487.36 | 602.23 | 1863.94 | -12119.23 |
| 7. | $Cd_6Se_5S_1$ | 615.17 | 66.62 | 1853.43 | -5222.52 |
| 8. | KDP | 49.18 | 85.30 | 855.023 | 17.731 |
| 9. | Urea | 38.22 | 42.48 | 122.61 | 14.49 |

All values are in a.u.

Figure 1 demonstrates a comparison of average $\beta$ all of stable geometries $Cd_1Se_1S_1$, $Cd_2Se_1S_1$, $Cd_3Se_2S_1$, $Cd_4Se_1S_3$, $Cd_4Se_3S_1$, $Cd_5Se_4S_1$, $Cd_6Se_5S_1$ against cluster size ($n$). It is observed that $\beta$ shows an increasing trend with an increase in cluster size, however certain anomalies are noticed for the cluster size of $n=2$ and $n=4$. Starting from $n=1$ the magnitude of $\beta$ reaches its maximum value at $n=4$ for cluster structure $Cd_4Se_1S_3$, whereas for the same value of $n$, $\beta$ shows a minimum value for $Cd_4Se_3S_1$. For the same order ($n$), $\beta$ being high as well as low illustrates the uniqueness of $Cd_4Se_mS_p$ cluster structure in the perspective of quantum optics. $\beta$ value is directly proportional to the reactivity of the probe system [58]. Larger $\beta$ means larger reactivity.

Similarly, the second order static hyperpoloarizability ($\gamma$) values of different energetically stable $Cd_nSe_mS_p$ geometries for clusters of size $n=1–6$ are shown in Figure 1 (Table 2). $\gamma$ also shows an increasing trend with larger cluster size, however, its nature is not uniform and a sudden jump is noticed at $n=4$ and 5. For a specific cluster structure $\beta$ and $\gamma$ values show opposite nature.

Figure 1: Mean first and second order hyper polarizability ($\beta$, $\gamma$) versus cluster size ($n$) graph of $Cd_1Se_1S_1$, $Cd_2Se_1S_1$, $Cd_3Se_2S_1$, $Cd_4Se_1S_3$, $Cd_4Se_3S_1$, $Cd_5Se_4S_1$ and $Cd_6Se_5S_1$ clusters in water.

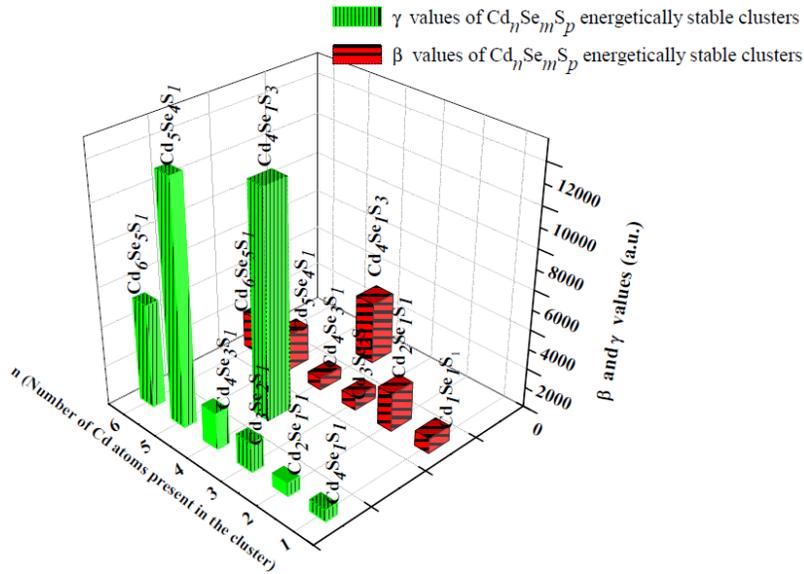

Both positive and negative values of $\beta$ and $\gamma$ establish the outstanding nonlinear property of $Cd_nSe_mS_p$ clusters, as a self-focusing or self-trapping and self-defocusing material [53] which are related to the nonlinear optical process induced by the change of refractive index of materials [59]. So $Cd_nSe_mS_p$ clusters can be considered as the good candidate for the designing of a new generation of highly efficient NLO active devices. Higher values of $\mu$, $\alpha$, $\beta$ and $\gamma$ of $Cd_nSe_mS_p$ clusters then popularly used NLO active materials [potassium dihydrogen phosphate (KDP) and Urea] justifies the potential application of $Cd_nSe_mS_p$ clusters as NLO active materials (Table 2). To provide an explanation for the unusual behavior on the basis of $\beta$ and $\gamma$ values, we have tried to analyze the Mulliken charges on Cd, Se and S atoms on each $Cd_nSe_mS_p$ cluster under investigation. SD 5 shows the variation of Mulliken charge per Cd atom against cluster size. A gradual decrease in the minimum charge of Cd atom is observed with an increase of cluster size except at $n=4$ and 5, where they increase. This unusual aspect of charge variation might be responsible for the sudden increase in static $\gamma$ in $Cd_nSe_mS_p$ cluster units for $n=4$ and 5. For larger cluster size decreasing tendency of charge on Cd atom is observed at $n=7$ and more.

Further, this unusual behavior can be verified by the molecular level explanation by viewing the gross population of HOMO-LUMO molecular orbitals for all energetically stable clusters $Cd_1Se_1S_1$, $Cd_2Se_1S_1$, $Cd_3Se_2S_1$, $Cd_4Se_3S_1$, $Cd_4Se_1S_3$, $Cd_5Se_4S_1$, $Cd_6Se_5S_1$ (Scheme 2). From a close view of molecular orbital, we found an evident localization in HOMO of the $Cd_2Se_1S_1$ $Cd_4Se_3S_1$, $Cd_5Se_4S_1$ and $Cd_4Se_1S_3$ clusters, however, their LUMO possess high degree of delocalization. Hence, from the localization and delocalization in HOMO and LUMO we can conclude that there is significant chance of some charge transfer interaction, which consequences in the drastic increment and decrement of value of $\beta$ and $\gamma$.

Scheme 2: HOMO, LUMO of $Cd_1Se_1S_1$, $Cd_2Se_1S_1$, $Cd_3Se_2S_1$, $Cd_4Se_1S_3$, $Cd_4Se_3S_1$, $Cd_5Se_4S_1$ and $Cd_6Se_5S_1$ clusters.

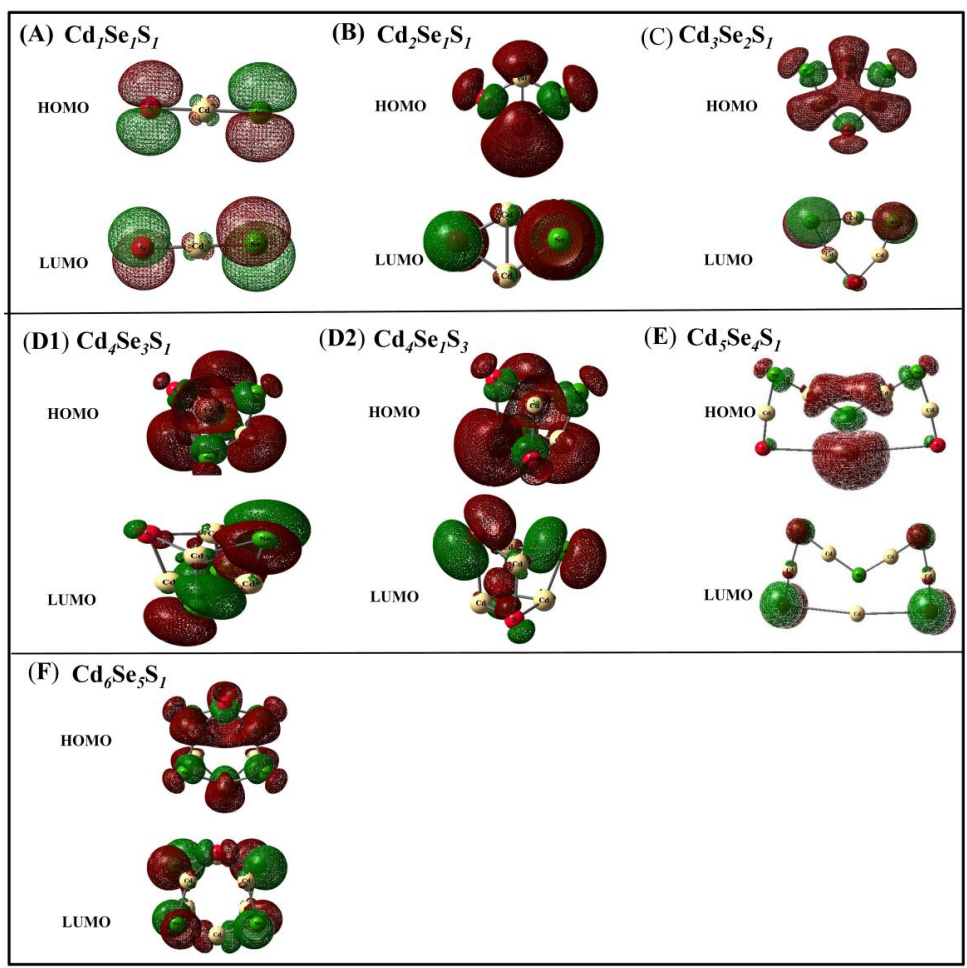

## 3.3 Optical absorption and emission of $Cd_nSe_mS_p$ clusters in water

Figure 2 and SD 6 describe the comparison between experimental and computed optical absorption spectra of different $Cd_nSe_mS_p$ clusters in water. For clusters in water medium the computed absorption peak with maximum oscillator strength ($f$) is observed to be red shifted with increasing the cluster size except $n=1$. The red shift in CdSe clusters has been reported by many authors [60]. So we can mention that optical absorption of CdSeS cluster is also size sensitive. But in some cases the absorption wavelength of large sized clusters is also observed to be shorter than those of small sizes. In the present section, we have compared the theoretical and experimental optical absorption and emission data of different $Cd_nSe_mS_p$ clusters. Absorption data for all suggested $Cd_nSe_mS_p$ clusters show the existence of very closely separated oscillator strengths (≤3meV) in all computed, which validates the criterian of dark and bright exciton interaction and HOMO-LUMO transitions in the semiconductor cluster [61]. For $n=4$ $Cd_4Se_1S_3$ cluster shows absorption bands for $S_0 \rightarrow S_1$, $S_0 \rightarrow S_2$ and $S_0 \rightarrow S_3$ transitions at 467.66 nm, 462.69 nm and 407.95 nm with a maximum absorbance for $S_0 \rightarrow S_2$ transition. Whereas, $Cd_4Se_1S_3$ shows the absorption bands at 420.10 nm, 414.18 nm and 411.61 nm with maximum absorbance for $S_0 \rightarrow S_2$ transition (not shown). Due to its high polarity and reactivity in presence of water environment a red shift in the band position is expected for $Cd_4Se_1S_3$ cluster. Experimentally CdSeS QD shows a strong absorption band at ~450 nm in a water medium. The $\Delta\varepsilon$ of CdSeS is obtained from an experimental spectrum is about 3.34 eV which matches with the computed $\Delta\varepsilon$ of $Cd_4Se_1S_3$ (3.56 eV). Similarly, we have compared the luminescence data both experimental and computed CdSeS QD structures (Figure 3). Experimentally, CdSeS QD shows a strong emission at ~500 nm in water medium for $S_1 \rightarrow S_0$ transition which matches with computed emission at 474 nm for $Cd_4Se_1S_3$. Experimental and computed absorption and emission data suggest that the $Cd_4Se_1S_3$ form of the CdSeS cluster will sustain in water medium due to its high polarity and reactivity. We have also calculated the average particle size (2R) of CdSeS clusters using the Henglein's empirical formula [62]:

$$2R_{(CdSeS)} = \frac{0.1}{(0.138 - 0.0002345\lambda_C)} \quad\ldots\ldots\ldots\ldots\ldots\ldots\ldots\ldots\ldots\ldots\ldots\ldots\ldots\ldots\ldots\ldots\ldots\ldots (7)$$

where $\lambda_C$ is absorption edge and 2R is the diameter of the QDs. The absorption edge ($\lambda_C$) can be easily determined by the intersection of sharply decreasing region of the spectrum with the baseline.

The average values of particle size have been calculated with the help of experimental and computed absorption spectra of different $Cd_nSe_mS_p$ clusters in water. The average values of particle size have been compared between experimental and computed CdSeS QD structures. The particle size of the experimentally prepared CdSeS cluster has been estimated as 3.9 nm for $\lambda_c$=280 nm (Figure 2), which closely matches with the particle size calculated for $Cd_4Se_3S_1$ cluster (3.5 nm for $\lambda_c$=266.67 nm). Thus, the particle size of prepared CdSeS cluster (3.9 nm) and computed $Cd_4Se_1S_3$ cluster (3.5 nm) establish a perfect correlation between experimental and computed results.

Figure 2: Electronic absorption of experimental CdSeS quantum dots and computed $Cd_4Se_1S_3$ cluster.

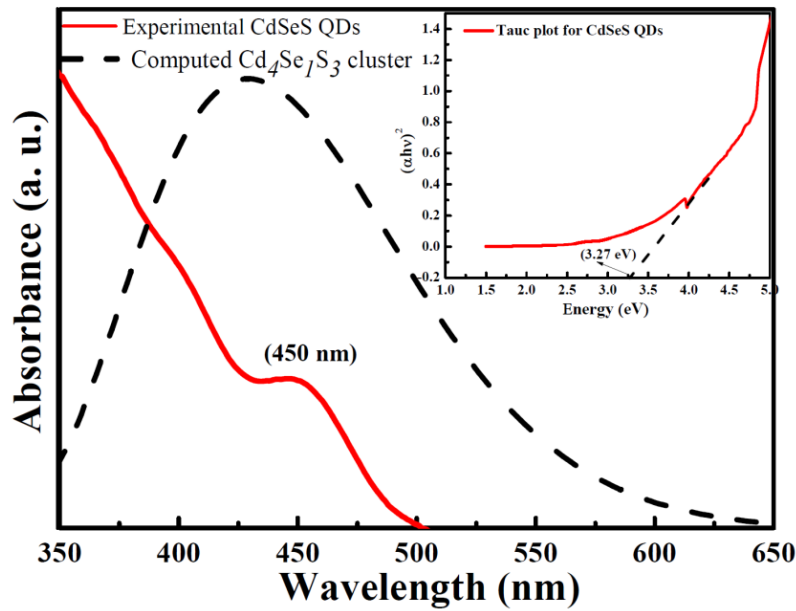

The compositions of Cd, Se and S in $Cd_nSe_mS_p$ QDs play a major role for the change in band gap and particle size [38]. Different Cd/Se/S feed molar ratio results change in the absorption, emission bandgap energy and size of CdSeS QDs. CdSe clusters have homogeneous structure, while CdSeS shows graded structure. Graded QDs show a variation in size and shape of cluster structure which is tuned by controlling the composition. The computed sizes of different $Cd_4Se_mS_p$ clusters indicate an increase in size with an increase in the Se feed amount. Such increment in size is consistent with the fact that the bond length of Cd-Se larger than that of Cd-S. Due to this asymmetric bond length variation in between Cd and Se, S the $Cd_nSe_mS_p$ structure becomes graded instead of homogeneous like CdSe or CdS. Also we have observed that the size of the $Cd_nSe_mS_p$ QDs depends on band gap and compositions. Lower band gap and more selenium result in larger size of QDs, thus resulting red shifted absorption and emission.

Figure 3: Fluorescence emission of experimental CdSeS Quantum dots and computed $Cd_4Se_1S_3$ cluster.

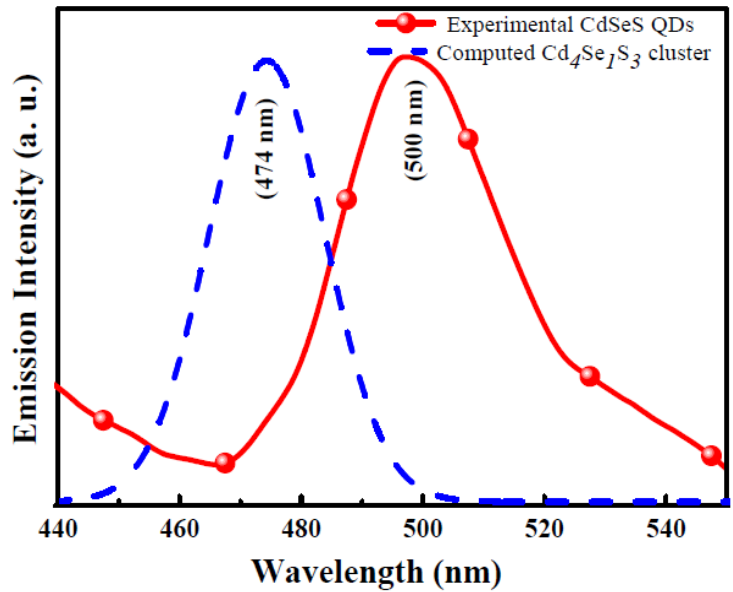

## 4. Conclusion

In this work we have reported accurate analyses of structural and NLO properties of $Cd_nSe_mS_p$ clusters up to $n=6$ for the first time. A detailed ab initio study on the energy, polarity, entropy, polarizability, hyperpolarizability, absorption, emission frequencies, etc. of these CdSeS clusters have been performed employing DFT approximation. Revealed trends suggest that the response properties are highly structure-dependent and is particularly noticeable in higher cluster. Clusters with same configuration with nonplanar distorted geometry exhibits higher stability and reactivity compared to clusters having planar structures. Energies of $Cd_nSe_mS_p$ clusters also reveal that for a fixed number of Cd atoms as the number of selenium atom increases than Sulphur atom, the stability of the cluster structure increases. $Cd_4Se_1S_3$ cluster establishes its higher reactivity due to smaller $\Delta\varepsilon$, larger $\mu$ and larger S values than the others in a water medium. $\Delta_2E$, first and second order hyperpolarizability values ($\beta$, $\gamma$) also validate the higher reactivity of $Cd_4Se_1S_3$. There are possibilities that the clusters $Cd_3Se_2S_1$, $Cd_4Se_3S_1$, $Cd_4Se_1S_3$, $Cd_5Se_4S_1$ and $Cd_6Se_5S_1$ can also exist in water medium due to the existence of similar computed physical parameters. The close correlation between experimental and computed electronic absorption and emission data validates the conclusion that in water medium the possible existence of cluster structure is $Cd_4Se_1S_3$. Symmetrized orbital analysis and Mulliken charge variation also provide the explanation of observed higher hyperpolarizability variation in CdSeS compared to the mostly available CdX. In accordance with all other reported measurements our computed and experimental results show that CdSeS clusters are far more polarizable and reactive than the mostly available CdS, CdSe and CdTe clusters.